\documentclass[fleqn,twoside]{article}
\usepackage{espcrc2}

\usepackage{graphicx}
\usepackage[figuresright]{rotating}

\newcommand{\AmS}{{\protect\the\textfont2
  A\kern-.1667em\lower.5ex\hbox{M}\kern-.125emS}}

\title{Effective string picture for confinement at finite
       temperature: theoretical predictions and high precision numerical
       results}

\author{M. Caselle\address[torino]{Universit\`a di Torino and INFN, Sezione di Torino,
        Via P. Giuria 1, I--10125 Torino, Italy, \\
        E-mail: caselle@to.infn.it, panero@to.infn.it},%
 M. Hasenbusch\address[nicdesy]{NIC/DESY Zeuthen,
 Platanenallee 6, D--15738 Zeuthen, Germany, \\
        E-mail: Martin.Hasenbusch@desy.de}
        and
        M. Panero\addressmark[torino]}

\begin{document}

\begin{abstract}
The effective string picture of confinement is used to derive
theoretical predictions for the interquark potential at finite
temperature. At short distances, the leading string correction to
the linear confining potential between a heavy quark-antiquark
pair is the ``L\"uscher term''. We assume a Nambu--Goto effective
string action, and work out subleading contributions in an
analytical way. We discuss the contribution given by a possible
``boundary term'' in the effective action, comparing these
predictions with results from simulations of lattice
$\mathbf{Z}_2$ gauge theory in three dimensions, obtained with an
algorithm which exploits the duality of the $\mathbf{Z}_2$ gauge model with the
Ising spin model. \vspace{1pc}
\end{abstract}

\maketitle

\section{EFFECTIVE STRING PICTURE}

The effective string picture is expected to provide a good
physical description of confining gauge theories in the low energy
regime.
Assuming that two confined color charges  
are joined by a thin flux tube that fluctuates like
a vibrating string, and describing the string world sheet dynamics
by an effective action $S_{\mbox{\tiny{eff}}}$,
one can derive quantitative predictions about the potential
between a confined quark--antiquark pair.
In particular, for
\emph{massless} string fluctuations, the simplest choice is the
Nambu--Goto string action: $ S_{\mbox{\tiny{eff}}} = \sigma \cdot
\mathcal{A} $, which is proportional to the area $\mathcal{A}$ of the
world--sheet surface and $\sigma$ is the \emph{string tension},
appearing as a parameter of the effective theory.

We consider a 3D system with the extension $L_s^2 \times L$  with
$L_s >> L$. With periodic boundary conditions employed (at least)
in the short direction,
the temperature $T$ is proportional to $1/L$.

Taking into account leading quantum fluctuations,
the result for the expectation value of the Polyakov loop
correlator is given by
\begin{equation}
\label{lo}
\langle P^\dagger (R) P(0) \rangle = \frac{ e^{-\sigma RL + k}
}{\eta \left( i\frac{L}{2R} \right)} \;,
\end{equation}
where $\eta$ is Dedekind's function. The term associated with the
minimal world sheet surface induces the exponential area-law fall
off, and the consequent linear rise in the interquark potential
$V(R)$. The first non-trivial contribution in
$S_{\mbox{\tiny{eff}}}$ results in the determinant of the Laplace
operator, and the corresponding contribution to the interquark
potential $V(R)$ --- in a regime of distances shorter than
$\frac{L}{2}$ --- is the L\"uscher term \cite{lsw}:
\begin{equation}
\label{potential} V(R)=-\frac{1}{L} \ln \langle P^\dagger (R) P(0)
\rangle \simeq \sigma R -\frac{\pi}{24 R} \;.
\end{equation}

Further terms in the expansion of the Nambu--Goto action give rise
to a contribution involving a combination of Eisenstein functions
\cite{eisen}:
\begin{equation}
\label{nlo} -\frac{\pi^2}{1152 \sigma R^3} \left[2E_4 \left(
i\frac{L}{2R} \right) -E_2^2 \left( i\frac{L}{2R} \right)\right] \;.
\end{equation}

However, it is worth mentioning that, while Eq. (\ref{lo}) and Eq.
(\ref{potential}) are generally accepted, on the other hand the
further contribution to $V(R)$ expressed by Eq. (\ref{nlo}) is
still under debate.

A possible ``boundary term'' in the effective action, as proposed in ref.
\cite{su3}, can be
treated by means of a perturbative expansion in $b$ (a parameter
proportional to the coefficient of such a term), which induces a
leading order correction like \cite{noi}:
\begin{equation}
R \longrightarrow \frac{R}{\sqrt{1+ \frac{2b}{R}}} \;,
\end{equation}
with a short distance contribution to $V(R)$
: $-\frac{b
\pi }{24 R^2}$.

\section{$\mathbf{Z}_2$ LATTICE GAUGE THEORY}
We run numerical simulations of $\mathbf{Z}_2$ lattice gauge
theory in three space-time dimensions \cite{noi}. We chose this
model because the effective string picture is believed to be
independent of the underlying gauge group, and $\mathbf{Z}_2$
gauge group is interesting from the point of view of the
\emph{center relevance} in confinement.
Moreover, the duality with the Ising spin model and a novel
algorithm allow us to reach high precision results.

The pure 3D lattice gauge model is described in terms of
$\sigma_{x,\mu} \in \mathbf{Z}_2$ variables defined on the lattice
bonds. The dynamics is governed by the standard Wilson action,
which enjoys $\mathbf{Z}_2$ gauge invariance. The partition
function is expressed in terms of plaquette variables $\sigma_{p}$
and reads:
\begin{equation}
Z(\beta)= \sum_{\mbox{c.} } e^{ -\beta S } = \sum_{\mbox{c.} }
\exp \left[ +\beta \sum_{p} \sigma_{p} \right] \;.
\end{equation}
There are different phases: a confined, strong coupling phase,
with massive string fluctuations for $\beta < 0.47542(1)$ \cite{HaPi}, a
confined, \emph{rough} phase, with massless string fluctuations
(this is the regime we studied in our simulations) and a deconfined
phase for 
$\beta > 0.7614134(2)$
\cite{BlTa}.

This model is \emph{dual} with respect to the $\mathbf{Z}_2$ Ising
spin model in 3D. The ratio of Polyakov-loop correlators $G(R)$ in the gauge
model can be expressed as a ratio of partition functions $Z_{L \times R}$
in the spin model:
\begin{equation}
\frac{G(R)}{G(R+1)} = \frac{Z_{L \times R } }{Z_{L \times (R+1) }} \;.
\end{equation}
The index $L \times R$ means that links perpendicular to an $L \times R$ plane
have an antiferromagnetic coupling constant. The ratio of partition functions
can be expressed as an expectation value in one of the two ensembles.
However the corresponding observable shows extremely large fluctuations.
This problem can be overcome by a factorization
\begin{equation}
\frac{Z_{L \times R } }{Z_{L \times (R+1) }}
= \prod_{i=0}^{L-1} \frac{Z_{[L \times R] +i} }{Z_{[L \times R]
+i+1} } \;,
\end{equation}
where  $Z_{[L \times R] +i+1}$ has just one more antiferromagnetic link
than $Z_{[L \times R] +i}$. The expectation values corresponding to these
ratios can be easily obtained. A similar method (\emph{snake algorithm})
has been employed in refs. \cite{snake} to compute the 't Hooft loop in
the $SU(2)$ gauge model.


Important features of our implementation are multi-level updating and
a hierarchical organization of sublattices. The CPU time is
roughly proportional to the inverse temperature $L$, and
\emph{independent of the distance $R$ between the quark sources},
thus the algorithm is particularly useful for large interquark
distances.

\section{NUMERICAL RESULTS}

Let $F(R,L)$ be the free energy of a heavy quark-antiquark pair at
finite temperature: $ G(R)=e^{-F(R,L)} $. We studied ``quantum
terms'' in free energy differences, by measuring the quantity
defined as:
\begin{equation}
Q(R,L) = F(R+1,L) - F(R,L) - \sigma L \;.
\end{equation}

\begin{figure}[htb]
\vspace{9pt}
\includegraphics[scale=0.55]{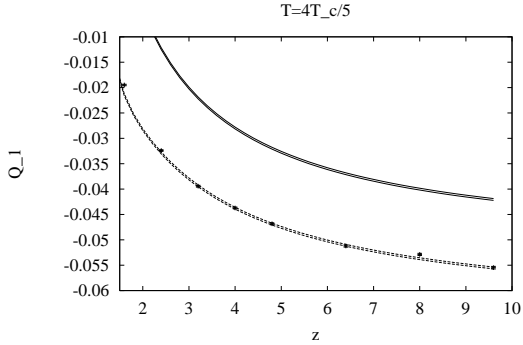}
\caption{$Q(R,L)$ for $L=10$ (i.e. $T=4T_c/5$) at $\beta=0.75180$.
$z$ is defined as: $z\equiv \frac{2R}{L}$. Solid curves correspond
to the free bosonic string prediction, while dashed lines are the
NLO Nambu--Goto correction. A pure area--law would be described by
$Q=0$.} \label{45tc}
\end{figure}

Fig. \ref{45tc} shows that for large interquark distances (namely:
$L<2R$) our numerical results are in good agreement with the NLO
prediction eq.~(\ref{nlo})
from Nambu--Goto string. It is important to note that
the agreement between numerical results and the NLO prediction is
not the result of a fitting procedure. A ``purely classical'' area
law is definitely ruled out, and the LO term alone is not
sufficient to describe the data. We also found that the
coefficient for a possible ``boundary term'' in the effective
action for this model is very small, suggesting that it is exactly
zero, i.e. $b=0$.

\begin{figure}[htb] \vspace{9pt}
\includegraphics[scale=0.55]{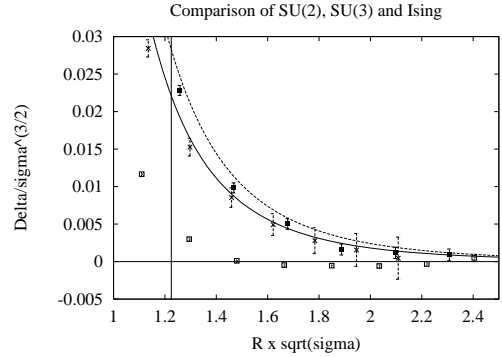}
\caption{Behaviour of $SU(3)$ \cite{su3} (white squares), $SU(2)$
\cite{su2} (crosses), and $\mathbf{Z}_2$ \cite{noi} (black
squares) gauge models at short distances. \label{newdelta_zoom} }
\end{figure}
Fig. \ref{newdelta_zoom} --- see \cite{noi} for details --- shows the
deviation of a quantity related to the free energy second
derivative from the free string
behaviour in the $L>2R$ regime. The normalization is chosen so as
to allow a meaningful comparison among different LGT's. Our
$\mathbf{Z}_2$ results and $SU(2)$ data are in good agreement:
this may be a possible signature of the center relevance to
confinement.

\section{CONCLUSIONS}

We tested the effective string predictions with precise numerical
data for the finite temperature $\mathbf{Z}_2$ lattice gauge
theory, using an algorithm that exploits the duality properties of
the model. We explored a wide range of distances and detected
next-to-leading order effects. Our numerical results seem to rule
out a possible ``boundary term'' in the effective string action
describing the present model.
For large distances $R$ and high temperatures (i.e. small $L$) we
see an excellent agreement of the numerical data with the NLO
prediction Eq.~(\ref{nlo}). In the regime of short distance and
low temperature, we compared our results with different gauge
theories, 
which have been studied by other authors. We focused onto the
deviation of a well-suited combination of free energy differences
with respect to the free string prediction. In particular, the
agreement between our $\mathbf{Z}_2$ results and $SU(2)$ data
might be interpreted as a possible signature of the fact that the
center degrees of freedom play an important role in the
confinement mechanism.

\end{document}